\documentclass[journal]{IEEEtran}
\usepackage{cite}

\ifCLASSINFOpdf
\else
\fi

\usepackage{stfloats}

\usepackage{amsmath,amssymb,amsfonts}
\usepackage{tabularx}
\usepackage{algorithm}
\usepackage{algpseudocode}
\usepackage{multicol}
\usepackage{multirow}
\usepackage{dirtytalk}
\usepackage{graphicx}
\usepackage{textcomp}
\usepackage{xcolor}
\usepackage{float}
\usepackage{lipsum}
\usepackage{subcaption}
\usepackage{comment}

\usepackage[english]{babel}
\usepackage{amsthm}
\usepackage{contour}
\usepackage{booktabs}
\usepackage{color,soul}

\usepackage{tikz}   
\usetikzlibrary{positioning}
\usetikzlibrary{patterns, arrows.meta}
\usetikzlibrary{shadows}
\usetikzlibrary{external}
\usetikzlibrary{calc}

\usetikzlibrary {shapes.misc}

\usepackage{xcolor}
\newcommand{\hlc}[2][yellow]{{%
    \colorlet{foo}{#1}%
    \sethlcolor{foo}\hl{#2}}%
}

\usepackage{pgf}
\usepackage{pgfplots}
\pgfplotsset{compat=1.18}
\usepgfplotslibrary{statistics}
\usepgfplotslibrary{fillbetween}

\usetikzlibrary{fit}

\usepackage[english]{babel}
\usepackage[nottoc]{tocbibind}
\def\BibTeX{{\rm B\kern-.05em{\sc i\kern-.025em b}\kern-.08em
    T\kern-.1667em\lower.7ex\hbox{E}\kern-.125emX}}

\usepackage[hidelinks]{hyperref}
\hypersetup{
  colorlinks,
  citecolor= blue,
  linkcolor=red,
  urlcolor=blue}

\begin{document}
%
\title{BEACON: A Bayesian Evolutionary Approach for Counterexample Generation of Control Systems}

\author{Joshua~Yancosek
        and~Ali~Baheri~
\thanks{Joshua~Yancosek is with the Department
of Mechanical, Materials and Aerospace Engineering, West Virginia University, Morgantown, WV 26506, USA. email: \tt\small jmy0004@mix.wvu.edu}
\thanks{Ali Baheri is with the Department
of Mechanical Engineering, Rochester Institute of Technology, Rochester,
NY 14623, USA. email: \tt\small akbeme@rit.edu}}

\maketitle

\begin{abstract}
The rigorous safety verification of control systems in critical applications is essential, given their increasing complexity and integration into everyday life. Simulation-based falsification approaches play a pivotal role in the safety verification of control systems, particularly within critical applications. These methods systematically explore the operational space of systems to identify configurations that result in violations of safety specifications. However, the effectiveness of traditional simulation-based falsification is frequently limited by the high dimensionality of the search space and the substantial computational resources required for exhaustive exploration. This paper presents BEACON, a novel framework that enhances the falsification process through a combination of Bayesian optimization and covariance matrix adaptation evolutionary strategy. By exploiting quantitative metrics to evaluate how closely a system adheres to safety specifications, \texttt{BEACON} advances the state-of-the-art in testing methodologies. It employs a model-based test point selection approach, designed to facilitate exploration across dynamically evolving search zones to efficiently uncover safety violations. Our findings demonstrate that \texttt{BEACON} not only locates a higher percentage of counterexamples compared to standalone BO but also achieves this with significantly fewer simulations than required by CMA-ES, highlighting its potential to optimize the verification process of control systems. This framework offers a promising direction for achieving thorough and resource-efficient safety evaluations, ensuring the reliability of control systems in critical applications. A Python implementation of the algorithm can be found at \url{https://github.com/SAILRIT/BO-CMA}.
\end{abstract}

\begin{IEEEkeywords}
Falsification, Bayesian optimization, covariance matrix adaptation evolutionary strategy, safety-critical systems
\end{IEEEkeywords}

\IEEEpeerreviewmaketitle

\section{Introduction}

\IEEEPARstart{T}{he} growth of cyber-physical systems, such as autonomous vehicles and robotics, has significantly raised the importance of ensuring controllers operate safely and reliably \cite{mitra2021verifying}. Traditional approaches to safety verification, such as formal verification, simulation-based testing, and model checking, each contribute valuable insights but also face significant limitations \cite{clarke1997model}. Formal verification methods, grounded in rigorous mathematical proofs, offer a high degree of assurance by proving system properties. However, their practical application is often constrained by the complex nature of control systems and the computational intensity required to analyze large state spaces—a phenomenon known as the state explosion problem. Model checking automates the process of verifying whether a system's model meets specified criteria \cite{plaku2013falsification}. While effective for discrete systems, model checking struggles with scalability and the complexities of systems with continuous states.

Simulation-based falsification, on the other hand, has become an essential aspect of safety validation in control systems, particularly in safety-critical systems \cite{kapinski2015simulation,baheri2023exploring,baheri2023safety}. Falsification pertains to the systematic discovery of counterexamples, or conditions under which the system fails to meet safety specifications. It serves as a tool in the design-time assurance process, especially when handling complex systems where conventional verification approaches fall short due to nonlinearities and high dimensionality. Software tools such as S-TaLiRo \cite{fainekos2012verification,annpureddy2011s}, Breach \cite{donze2010breach}, C2E2 \cite{duggirala2015c2e2}, and DryVR \cite{qi2018dryvr} have been instrumental in system falsification. These tools are designed to automate and facilitate the falsification process. They provide frameworks for systematically exploring the system's behavior under various conditions, seeking configurations that lead to specification violations. Various falsification methods, including search-based testing \cite{zhang2018two,ramezani2021testing,zhang2021effective,hekmatnejad2020search,zhang2020hybrid,deshmukh2015stochastic,ernst2021falsification}, optimization-based testing \cite{ramezani2022optimization,ramezani2020multiple,mathesen2021efficient,deshmukh2017testing,abbas2019safe,aerts2018temporal,annapureddy2010ant} and machine learning approaches \cite{akazaki2018falsification,zhang2019multi,qin2019automatic,beard2022safety} offer different strengths. 

Several works have demonstrated the potential of hybrid methodologies in the domain of safety verification. Some approaches have integrated symbolic methods, which provide strong guarantees on correctness and completeness, with numeric methods, known for their efficiency and scalability \cite{frehse2011spaceex}. The integration of machine learning algorithms with traditional search or optimization-based falsification methods has been explored to predict areas of the parameter space more likely to yield counterexamples \cite{dreossi2019compositional}. This predictive capability can guide the falsification process more effectively, reducing the number of simulations required.  Approaches that adaptively adjust their search have shown promise \cite{koren2018adaptive,bartocci2021adaptive}. 

Despite these advancements, a significant gap remains in the ability to efficiently identify safety violations in complex control systems. Many of the existing hybrid approaches still face challenges in balancing exploration and exploitation, dealing with high-dimensional spaces. We propose the \texttt{BEACON} framework, a \textbf{\ul{B}}ayesian \textbf{\ul{E}}volutionary \textbf{\ul{A}}pproach for \textbf{\ul{CO}}u\textbf{\ul{N}}terexample Generation. It is designed to tackle these challenges by integrating Bayesian optimization (BO) with covariance matrix adaptation evolutionary strategy (CMA-ES) into a cohesive hybrid strategy. The rationale behind this integration is twofold. At its core, BO excels in efficiently exploring search spaces by using a probabilistic model to guide the search process. This model-based approach enables BO to make informed decisions about where to sample next, optimizing the trade-off between exploration (searching in new areas) and exploitation (focusing on areas with known potential). However, BO's performance can be limited by the accuracy of the surrogate model and its tendency to focus too narrowly on regions of perceived interest, potentially overlooking other critical areas of the search space. CMA-ES addresses some of the limitations inherent in BO by employing an evolutionary strategy that adaptively refines the search based on the fitness of previous candidates. Its ability to dynamically adjust the search distribution (mean and covariance) based on evolutionary principles allows for a more flexible exploration of the search space. This adaptiveness is particularly advantageous in exploring problems with multiple local optima.

\noindent \textbf{Our Contributions.} In this paper, we present the following contributions:
\begin{itemize}
    \item We propose a novel framework that synergistically merges BO and CMA-ES to efficiently uncover counterexamples in complex, high-dimensional uncertainty spaces.

    \item We conduct an extensive evaluation of our framework, emphasizing its adaptability and effectiveness in refining the search strategy for optimal falsification results.

    \item We release the \texttt{BEACON} framework as open-source software to facilitate the adoption of our approach within the safety verification community.
    
    \end{itemize}

\noindent \textbf{Paper Organization.} Section \ref{Preliminaries} presents the foundational background to the proposed framework and the problem statement. Section \ref{Methodology} describes the \texttt{BEACON} framework. Section \ref{evaluation} presents our experimental setup, results, and discussion. Finally, Section \ref{Conclusion} concludes the paper with a summary of the findings and future works.

\section{Preliminaries and Problem Setup}
\label{Preliminaries}

\begin{table}[t]
\caption{Reference of symbols commonly used throughout the paper.}
\begin{tabularx}{\linewidth}{p{1cm} X} 
\toprule
\textbf{Symbol} & \textbf{Definition}\\
\midrule
$\varphi$ & Safety specification\\
$\rho_{\varphi}(.)$ & Robustness function\\
$\mathcal{U}_{G}$ & Global search space\\
$\mathcal{U}_{L}$ & Local search zone\\
$\mathbf{e}$ & A set of environmental parameters\\
$e_{n}$ & An environmental parameter\\
$S({\mathbf{e}})$ & Signal/trajectory associated with the corresponding set of environmental parameters\\
\midrule
$b$ & Simulation budget\\
$P$ & Simulation budget in a local parameter zone\\
$n$ & Number of environmental parameters\\
\bottomrule
\end{tabularx}
\label{Symbol Ref}
\end{table}

\label{SSS}
\subsection{Signal Temporal Logic}
\label{STL}
Specifications consist of properties (predicates) $\psi$ over a continuous time signal. These properties are expressed in the formal language of signal temporal logic (STL) \cite{maler2013monitoring}. STL formulas are defined by:
\begin{equation}
    \begin{aligned}
    \varphi := \top \mid  \psi \mid \neg \varphi \mid \varphi_{1} \wedge \varphi_{2} \mid 
    \varphi_{1} \vee \varphi_{2} \mid \\
    \mathbf{G}_{[a,b]} \varphi \mid \mathbf{F}_{[a,b]} \varphi \mid \varphi_{1} \mathcal{U}_{[a,b]} \varphi_{2} \label{eqA}
    \end{aligned}    
\end{equation}
Besides $\psi$ which stands for a predicate or requirement, each of the following symbols represents a Boolean operator. $\top$ denotes the \emph{true} operator; $\neg$ denotes \emph{negation}; $\wedge$ denotes \emph{and}; $\vee$ denotes \emph{or}. $\mathbf{G}_{[a,b]}$ denotes globally true (\emph{always}) over a time range $[a,b]$ where $a,b\in \mathbb{R}_{[0,\infty]}$ and $a\leq b$. $\mathbf{F}_{[a,b]}$ denotes \emph{eventually}. \emph{Eventually} states that the specification is true at some point within the time range. $\mathbf{U}_{[a,b]}$ denotes \emph{until} which states that $\varphi_{1}$ remains true until $\varphi_{2}$ has been met. 

\noindent \textbf{Quantitative STL Semantics.} STL formulas reveal whether a specification has been violated or not. By incorporating quantitative semantics, defined in Table \ref{Quant_Sem}, one acquires a measure of robustness, how well a signal follows the specification \cite{donze2010robust}. A positive robustness value indicates the specification is satisfied; a negative robustness value indicates the specification has been violated. 

\begin{table}[b]
\caption{Quantitative semantics}
\label{Quant_Sem}
\begin{tabularx}{\linewidth}{p{2.5cm} X} 
\toprule
$\rho (S_{\mathbf{e}}(t),\psi )$ & $= \phantom{\_} c-\psi (\mathbf{x}[t])$\\
$\rho (S_{\mathbf{e}}(t),\neg\varphi )$ & $= \phantom{\_} -\rho (S_{\mathbf{e}}(t),\varphi )$\\
$\rho (S_{\mathbf{e}}(t),\varphi_{1}\wedge\varphi_{2})$ & $= \phantom{\_} \text{min}(\rho (S_{\mathbf{e}}(t),\varphi_{1}),\rho (S_{\mathbf{e}}(t),\varphi_{2}))$\\
$\rho (S_{\mathbf{e}}(t),\varphi_{1}\vee\varphi_{2})$&$= \phantom{\_} \text{max}(\rho (S_{\mathbf{e}}(t),\varphi_{1}),\rho (S_{\mathbf{e}}(t),\varphi_{2}))$\\
$\rho (S_{\mathbf{e}}(t),\mathbf{G}_{[a,b]}\varphi)$&$= \phantom{\_} \text{min}_{{t}'\in[t+a,t+b]} \rho(S_{\mathbf{e}}({t}'),\varphi)$\\
$\rho (S_{\mathbf{e}}(t),\mathbf{F}_{[a,b]}\varphi)$&$= \phantom{\_} \text{max}_{{t}'\in[t+a,t+b]} \rho(S_{\mathbf{e}}({t}'),\varphi)$\\
$\rho (S_{\mathbf{e}}(t),\varphi_{1}\mathbf{U}\varphi_{2})$&$= \phantom{\_} \text{max}_{{t}'\in [t+a,t+b]} (\text{min}(\rho(S_{\mathbf{e}}({t}'),\varphi_2),$\\
& $\phantom{= \_}\text{min}_{{t}''\in [0,{t}']}\rho(S_{\mathbf{e}}({t}''),\varphi_{1})))$\\
\bottomrule
\end{tabularx}
\end{table}
\subsection{Bayesian Optimization}
\label{Bayesian Opt}

BO is a sample-efficient, global optimization technique for expensive-to-evaluate, black-box objective functions that lack explicit analytical forms or are non-convex and non-differentiable, widely applied across various domains including machine learning, robotics, control, and design optimization problems \cite{wu2019hyperparameter,berkenkamp2023bayesian,baheri2017real,baheri2017iterative,baheri2020waypoint}. BO has emerged as a tool for falsification tasks, where the goal is to discover system configurations that lead to undesirable behaviors or safety specification violations \cite{deshmukh2017testing,ghosh2018verifying,shahrooei2023falsification}. Given the unknown relationship between the robustness function and environment parameters, BO uses surrogate modeling, typically through Gaussian processes (GP), to approximate this function based on observed data.

Consider we have a set of $n$ observations from previously evaluated environmental parameters, represented as $\mathbf{y}_n=\left[\hat{\rho}_{\varphi}\left(\mathbf{e}_1\right), \ldots, \hat{\rho}_{\varphi}\left(\mathbf{e}_n\right)\right]$ at environmental parameters $\mathbf{e}_1, \ldots, \mathbf{e}_n$. Here, $\hat{\rho}_{\varphi}(\mathbf{e})=\rho_{\varphi}(\mathbf{e})+\omega$ incorporates Gaussian noise $\omega \sim \mathcal{N}\left(0, \sigma^2\right)$. The posterior distribution of $\rho_{\varphi}(\mathbf{e})$ is characterized by the following equations for the mean $m_n(\mathbf{e})$, covariance $k_n\left(\mathbf{e}, \mathbf{e}^{\prime}\right)$, and variance $\sigma_n(\mathbf{e})$:

\begin{equation}
\label{GP_mean}
    m_{n}(\mathbf{e})=\mathbf{k}_{n}(\mathbf{e})(\mathbf{K}_{n}+\mathbf{I}_{n}\sigma^{2})^{-1}\mathbf{y}_n
\end{equation}
\begin{equation}
\label{GP_Cov}
    k_{n}(\mathbf{e},\mathbf{e'})=k(\mathbf{e},\mathbf{e'})-\mathbf{k}_{n}(\mathbf{e})(\mathbf{K}_{n}+\mathbf{I}_{n}\sigma^{2})^{-1}\mathbf{k}_{n}^{T}(\mathbf{e'})
\end{equation}
\begin{equation}
\label{GP_Var}
    \sigma_{n}^{2}(\mathbf{e})=k_{n}(\mathbf{e},\mathbf{e'})
\end{equation}
%
The covariance between a new set of environmental parameters and the previous ones is captured in the vector $\mathbf{k}_n(\mathbf{e})=\left[k\left(\mathbf{e}, \mathbf{e}_1\right), \ldots, k\left(\mathbf{e}, \mathbf{e}_n\right)\right]$. Here, $\sigma_n^2(\mathbf{e})$ denotes the variance, $\mathbf{I}_n$ represents the identity matrix, and $\mathbf{K}_n$ refers to the kernel matrix with entries $\left[k_n\left(\mathbf{e}, \mathbf{e}^{\prime}\right)\right]$.


\subsection{Covariance Matrix Adaptation Evolutionary Strategy (CMA-ES)}
\label{sec:CMA-ES}
The covariance matrix adaptation evolutionary strategy (CMA-ES) is a powerful optimization algorithm that belongs to the class of evolutionary algorithms. It is particularly well-suited for solving high-dimensional, non-convex optimization problems \cite{hansen2016cma}. Central to CMA-ES is its strategy of exploiting the correlations among variables to steer the search towards the global optimum efficiently. This is achieved through a mechanism that dynamically adapts the search strategy based on the history of previous evaluations.

CMA-ES initiates its process by generating a population of $P$ candidate solutions $\mathbf{e}_{1},...,\mathbf{e}_{P}$, each representing a set of environmental parameters. These candidates are sampled from a multivariate normal distribution defined as:
\begin{equation}
    \mathbf{e}_{i} \sim \mathcal{N}(\boldsymbol{\mu},\lambda^{2}\mathbf{C}), \quad i=1,...,P
\end{equation}
where $\boldsymbol{\mu}$ represents the mean vector, $\mathbf{C}$ denotes the covariance matrix capturing the relationship between variables, and $\lambda$ signifies the scale or step size of the search. To refine its search strategy, CMA-ES evaluates the generated candidates based on the robustness function $\rho_{\varphi}(\mathbf{e})$ and ranks them according to their performance, from the least to the most robust outcomes:
\begin{equation}
    \rho_{\varphi}(\mathbf{e}_{1}) \leq ... \leq \rho_{\varphi}(\mathbf{e}_{P})
\end{equation}
Subsequently, the algorithm updates the mean $\boldsymbol{\mu}$ and the covariance matrix $\mathbf{C}$ based on the top-performing candidates, identified as $P_{best}$, where $P_{best} \leq P$. The updated mean and covariance matrix are calculated as follows:
\begin{equation}
    \boldsymbol{\mu} = \frac{1}{P_{best}}\sum_{i=1}^{P_{best}}\mathbf{e}_{i}
\end{equation}
\begin{equation}
    \sigma_{nn'}^{2} = \frac{1}{P_{best}}\sum_{i=1}^{P_{best}}(\mathbf{e}_{i,n}-\mu_{n})(\mathbf{e}_{i,n'}-\mu_{n'}),
\end{equation}
where $\sigma_{nn'}^{2}$ represents the elements of the covariance matrix $\mathbf{C}$, and $\mathbf{e}_{i,n}$ and $\mu_{n}$ are the $n^{th}$ components of the $i^{th}$ environmental parameter vector and the mean vector, respectively.

This adaptive process allows CMA-ES to iteratively refine its search distribution, progressively focusing on more promising regions of the search space. The algorithm continues this evolutionary cycle, adjusting $\boldsymbol{\mu}$ and $\mathbf{C}$ with each generation, until a predefined termination criterion is met.

\subsection{Problem Statement}

Given a system under test (SUT) embedded within a high-dimensional parameter space $\mathcal{U}_{G} \subseteq \mathbb{R}^n$, the goal of falsification is to identify sets of environmental parameters $\mathbf{e} \in \mathcal{U}_{G}$ that lead the SUT to violate one or more predefined safety specifications. These safety specifications are formalized as constraints over the system's output trajectories $S(\mathbf{e})$, where $S: \mathbb{R}^n \rightarrow \mathbb{R}^m$ is a function mapping environmental parameters to system responses. A safety specification $\varphi$ is inherently defined in relation to the trajectories of a SUT. We interpret $\varphi$ to include all finite-horizon trajectories $S(\mathbf{e})$ that adhere to the defined system safety requirements. A trajectory $S(\mathbf{e})$, resulting from a set of environmental parameters $\mathbf{e}$, is deemed compliant with the specification $\varphi$ if and only if $S(\mathbf{e}) \in \varphi$. This condition is denoted as $S \models \varphi$, meaning that $\varphi$ evaluates the trajectory $S(\mathbf{e})$ as satisfying the safety specification. The structure of $\varphi$ is derived from multiple individual conditions, termed predicates. These predicates act as the elemental logical units that, through a combination of logical operations, construct the overall safety specification. Each predicate $\mu$ is considered a continuous function evaluated along the trajectory $S(\mathbf{e})$. Satisfaction of a predicate occurs when $\mu(S(\mathbf{e})) > 0$, indicating adherence to the safety criterion; otherwise, the predicate—and by extension, the trajectory—is considered falsified. Instead of merely assessing the Boolean satisfaction of a predicate, the notion of robust or quantitative semantics is introduced to measure the extent of satisfaction \cite{donze2010robust}. This approach introduces a more refined safety assessment by associating a real-valued function $\rho_{\phi}(S(\mathbf{e}))$ with each predicate, which is evaluated along the system trajectory $S(\mathbf{e})$.  This function serves as a \say{measure} of how significantly the safety specification is satisfied. The falsification task can thus be represented as an optimization problem:
\begin{equation}
    \underset{\mathbf{e}}{\mathrm{argmin}}\, \rho _{\varphi }\left ( S \left ( \mathbf{e} \right ) \right ) \label{(1)}
\end{equation}
where $\rho_{\varphi}(S(\mathbf{e}))$ is a robustness metric quantifying the degree of safety specification violation by the system's output for a given set of parameters $\mathbf{e}$. 
 
The \texttt{BEACON} framework addresses this optimization problem through a hybrid strategy that combines the exploratory strengths of BO with the adaptive capabilities of CMA-ES. Specifically, the framework partitions the global search space $\mathcal{U}_{G}$ into localized search zones $\mathcal{U}_{L}$, each potentially containing parameter sets that lead to specification violations.

\noindent{\textbf{Bayesian Optimization Component.}} BO is applied within each $\mathcal{U}_{L}$ to efficiently identify parameter sets that are likely to violate safety specifications. This is achieved by constructing a probabilistic model (e.g., a GP) of the robustness metric $\rho_{\varphi}(S(\mathbf{e}))$ and using acquisition functions to guide the selection of new parameter sets for evaluation.

\noindent{\textbf{Covariance Matrix Adaptation Evolutionary Strategy (CMA-ES) Component.}} CMA-ES is used to adaptively refine the search within $\mathcal{U}_{L}$ based on the outcomes of previous evaluations. It adjusts the sampling distribution (mean and covariance) to concentrate future simulations in regions of the parameter space more likely to uncover falsifying examples.

\begin{figure*}[!t]
\centering
\includegraphics{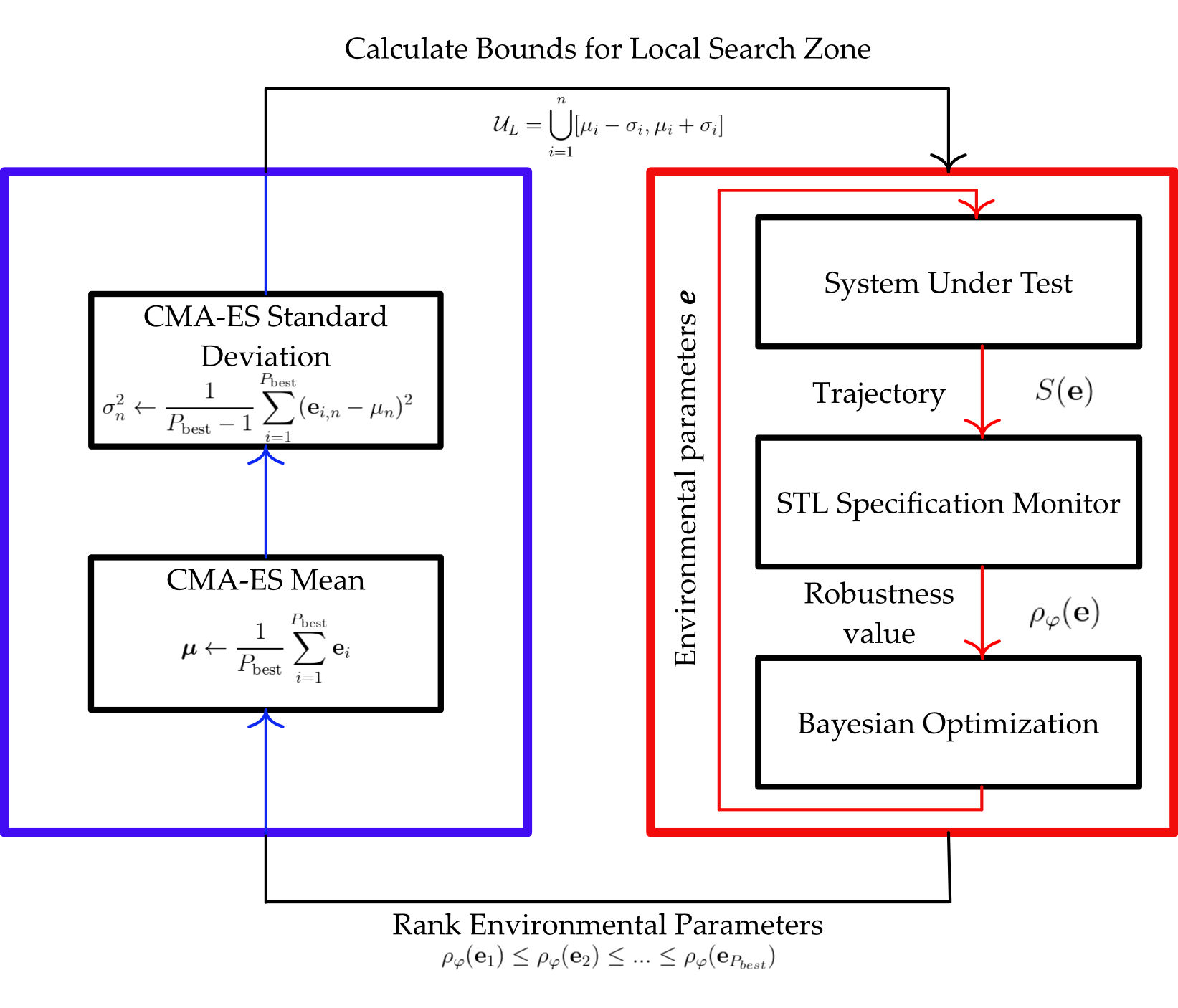}
\caption{Schematic Representation of the \texttt{BEACON} Falsification Framework. This framework constructs a model for evaluating system specifications within a defined local search zone, $\mathcal{U}_L \subseteq$ $\mathcal{U}_G$, as highlighted by the red box. Over $P$ iterations, BO is used to select new environmental parameters for simulation within $\mathcal{U}_L$. Upon exhausting the iteration budget $P$, the framework uses the $P_{\text {best }}$ environmental parameters to derive the mean and standard deviation, using principles from the CMA-ES, as indicated in blue. These statistical measures are then used to determine the upper and lower bounds of the subsequent local search zone, setting the stage for the next cycle of the process.}
\label{BOCMA Framework}
\end{figure*}   

\section{Methodology}
\label{Methodology}

This section details the \texttt{BEACON} framework, a novel approach that integrates BO and CMA-ES strategy to advance the falsification of control systems. The framework is designed to efficiently identify counterexamples in complex, high-dimensional search spaces characterized by numerous local optima. By synergistically combining the explorative capabilities of BO with the global search strategy of CMA-ES, \texttt{BEACON} aims to significantly reduce the number of simulations required to locate violations of safety specifications. The core strategy is depicted in Figure \ref{BOCMA Framework}.

The methodology uses a strategic division of the global search space, $\mathcal{U}_{G}$, into localized search zones, $\mathcal{U}_{L} \subseteq \mathcal{U}_{G}$, each defined by the adaptive mechanisms inherent in CMA-ES. This partitioning enables focused exploration and exploitation within subsets of the search space. Within each localized search zone $\mathcal{U}_{L}$, the BO constructs a GP model to serve as a surrogate for the system's robustness function, $\rho_{\varphi}(\mathbf{e})$.
To guide the selection of new test points within $\mathcal{U}_{L}$, the framework uses the lower confidence bound (LCB) acquisition function, balancing the exploration of unexplored regions against the exploitation of known areas of interest:
\begin{equation}
    \label{LCB_Eq}
    \mathbf{e}_{n}=\mathop{\mathrm{argmin}}_{\mathbf{e}\in\mathcal{U}_{L}} m_{n-1}(\mathbf{e})-\xi ^{\frac{1}{2}}\sigma^{BO}_{n-1} (\mathbf{e})
\end{equation}
where $\xi$ dynamically adjusts the focus between exploring new areas and exploiting existing knowledge to efficiently converge on global minima.

\begin{figure*}[b]
    \centering
 \begin{subfigure}[b]{0.3\textwidth}
    \centering
       \begin{tikzpicture}[scale=0.5]
    \begin{axis}[
           axis lines=left,
            axis line style={-},
            grid=both,
            xmin=0, xmax=20,
            ymin=0, ymax=20,
            xlabel={x},
            ylabel={y},
        ]
        \draw[red,line width=2pt](axis cs:1,1) -- (axis cs:1,17);
        \draw[red,line width=2pt](axis cs:1,17) -- (axis cs:16,17);
        \draw[red,line width=2pt](axis cs:16,17) -- (axis cs:16,1);
        \draw[red,line width=2pt](axis cs:1,1) -- (axis cs:16,1);
         \addplot[mark=*, only marks, color=blue] coordinates {
            (2,14)
            (13,12.4)
            (4.4,3.7)
            (8,4.4)
            (13.3,11.5)
        };

        \end{axis}
    \end{tikzpicture}
    \caption{First Local Search Zone}
  \end{subfigure}
  \hfill
   \centering
 \begin{subfigure}[b]{0.3\textwidth}
    \centering
       \begin{tikzpicture}[scale=0.5]
    \begin{axis}[
           axis lines=left,
            axis line style={-},
            grid=both,
            xmin=0, xmax=20,
            ymin=0, ymax=20,
            xlabel={x},
            ylabel={y},
        ]
        \draw[blue,line width=2pt](axis cs:3.63,4.92) -- (axis cs:3.63,13.48);
        \draw[blue,line width=2pt](axis cs:3.63,13.48) -- (axis cs:12.65,13.48);
        \draw[blue,line width=2pt](axis cs:12.65,13.48) -- (axis cs:12.65,4.92);
        \draw[blue,line width=2pt](axis cs:3.63,4.92) -- (axis cs:12.65,4.92);
         \addplot[mark=*, only marks, color=orange] coordinates {
            (7.2,9.6)
            (5,8)
            (6,11.7)
            (8,5.2)
            (7.6,9.5)
        };

        \end{axis}
    \end{tikzpicture}
    \caption{Second Local Search Zone}
  \end{subfigure}
    \hfill
   \centering
 \begin{subfigure}[b]{0.3\textwidth}
    \centering
       \begin{tikzpicture}[scale=0.5]
    \begin{axis}[
           axis lines=left,
            axis line style={-},
            grid=both,
            xmin=0, xmax=20,
            ymin=0, ymax=20,
            xlabel={x},
            ylabel={y},
        ]
        \draw[orange,line width=2pt](axis cs:5.65,6.65) -- (axis cs:5.65,10.95);
        \draw[orange,line width=2pt](axis cs:5.65,10.95) -- (axis cs:7.87,10.95);
        \draw[orange,line width=2pt](axis cs:7.87,10.95) -- (axis cs:7.87,6.65);
        \draw[orange,line width=2pt](axis cs:5.65,6.65) -- (axis cs:7.87,6.65);
         \addplot[mark=*, only marks, color=black] coordinates {
            (6.1,8.7)
            (5.8,8.4)
            (7.2,9.7)
            (6.6,8.2)
            (7.6,8.5)
        };
        
        \end{axis}
    \end{tikzpicture}
    \caption{Third Local Search Zone}
  \end{subfigure}
    \caption{Illustration of the \texttt{BEACON} methodology applied within a $2$-dimensional global search space $\mathcal{U}_{G}=[0,20]^2$. Each subfigure shows the evolving boundaries of local search zones, with the highlighted points representing the $P_{best}$ environmental parameters selected to refine the subsequent search space. This sequential adaptation showcases the framework's progression through the search space to efficiently explore regions of interest.}
    \label{fig:enter-label}
\end{figure*}
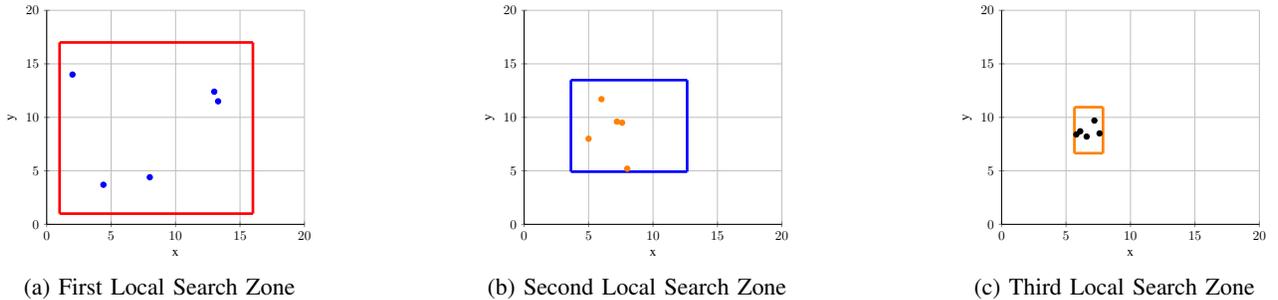

After simulating a set of $P$ environmental parameters within the current localized search zone $\mathcal{U}_{L}$, the framework applies CMA-ES's adaptive mechanisms to update the search strategy. This process begins by evaluating the robustness of the $P_{\text{best}}$ performing parameters, which then informs the calculation of the mean vector, $\mu$, and the variance, $\sigma^2$. These statistical parameters are crucial for shaping the boundaries of the next local search zones:
\begin{equation}
    \mathcal{U}_{L}=\bigcup_{i=1}^{n}[\mu_{i}-\sigma_{i},\mu_{i}+\sigma_{i}]
    \label{eqn:union}
\end{equation}
The Eq. \ref{eqn:union} shows how the \texttt{BEACON} framework dynamically tailors local search zones, $\mathcal{U}_{L}$, through a union of intervals across each dimension of the input space. Each interval is centered around the mean, $\mu_i$, of the best-performing parameters, expanded by their standard deviation, $\sigma_i$. This adjustment ensures that the search zones are not only concentrated around the most informative regions identified thus far but also sufficiently broad to explore areas that may harbor undiscovered counterexamples. Figure \ref{fig:enter-label} visualizes the evolution of $3$ consecutive search zones in a $2$-dimensional global search space. Each subsequent search space contracts around the $P_{best}$ environmental parameters highlighted in the prior search zone.

In the iterative exploration of local search zones by \texttt{BEACON}, we closely monitor the evolution of robustness values to identify any signs of stagnation. Stagnation occurs when the algorithm fails to find lower robustness values in successive local search zones, indicating a potential trap in a local optimum or having reached the vicinity of the global minimum. Mathematically, this is evaluated by comparing the minimum robustness values between consecutive search zones, as follows. Let $\mathbf{\hat{z}} = [\mathbf{e}_{1}^{(g)}, \ldots, \mathbf{e}_{P}^{(g)}]$ denote the set of environmental parameters evaluated in the current search zone ($g$), and $\mathbf{\hat{y}} = [\mathbf{e}_{1}^{(g-1)}, \ldots, \mathbf{e}_{P}^{(g-1)}]$ represent those from the previous zone ($g-1$). The corresponding robustness vectors for these sets are $\mathbf{z} = [\rho_{\varphi}(\mathbf{e}_{1}^{(g)}), \ldots, \rho_{\varphi}(\mathbf{e}_{P}^{(g)})]$ and $\mathbf{y} = [\rho_{\varphi}(\mathbf{e}_{1}^{(g-1)}), \ldots, \rho_{\varphi}(\mathbf{e}_{P}^{(g-1)})]$, respectively. Stagnation is formally detected when:
\begin{equation}
    \mathrm{min}(\mathbf{z}) > \mathrm{min}(\mathbf{y})
\end{equation}
implying no improvement in robustness has been achieved in the most recent search zone compared to its predecessor.

To address potential stagnation and avoid exhaustive focus on suboptimal regions, the \texttt{BEACON} framework incorporates a stagnation monitoring mechanism. This mechanism activates when the search fails to yield improved robustness outcomes over successive iterations. A predetermined threshold, $\delta$, represents the maximum number of consecutive local search zones allowed without observing any improvement. The counter, $\gamma$, tracks the number of such zones, and once $\gamma$ equals or exceeds $\delta$, it triggers a shift. This shift entails relocating the search focus to an unexplored area of the global search space, $\mathcal{U}_{G}$, in pursuit of new counterexamples.

We present the \texttt{BEACON} framework in Algorithm \ref{BO_CMA_Alg}. The algorithm initiates with a set of random samples for simulation and records the associated robustness values. Using equations from CMA-ES, the mean and variance are calculated to determine the boundaries of the subsequent local search space (line $6-8$). Within this new local search space, BO selects environmental parameters for simulation based on the LCB acquisition function (line $11$). After simulating the chosen parameters, a GP model is constructed to map the inputs to their robustness values (line $12$). If stagnation is observed, the search is redirected to a new region (line $16-23$).
\begin{algorithm}[t]
    \caption{\texttt{BEACON}: \textbf{B}ayesian \textbf{E}volutionary \textbf{A}pproach for \textbf{CO}u\textbf{N}terexample Generation}
    \label{BO_CMA_Alg}
    \begin{algorithmic}[1] 
        \State \textbf{Input:} global search space $\mathcal{U}_{G}$, global simulation budget $b$, search zone simulation budget $P$, and stagnation factor $\delta$
        \State \textbf{Initialize:} simulate random samples 
        \State \textbf{Until} simulation count $=$ b \textbf{do}:
        \For {$i = 1$ \textbf{to} $n$}
            \State $\boldsymbol{\mu} \gets \frac{1}{P_{\text{best}}}\sum_{i=1}^{P_{\text{best}}}\mathbf{e}_{i}$ \Comment{Calculate CMA-ES mean}
            \State $\sigma_{n}^{2} \gets \frac{1}{P_{\text{best}}-1}\sum_{i=1}^{P_{\text{best}}}(\mathbf{e}_{i,n} - \mu_{n})^{2}$ \Comment{Calculate CMA-ES variance}
            \State $\mathcal{U}_{L} \gets \bigcup_{i=1}^{n}[\mu_{i} - \sigma_{i}, \mu_{i} + \sigma_{i}]$  
        \EndFor
        \State Initialize local GP model of $\mathcal{U}_{L}$ 
        \For {$p = 1$ \textbf{to} $P$}
            \State $\mathbf{e}_{n} \gets \mathop{\mathrm{argmin}}_{\mathbf{e}\in\mathcal{U}_{L}} m_{n-1}(\mathbf{e}) - \xi^{\frac{1}{2}}\sigma^{BO}_{n-1} (\mathbf{e})$ \Comment{Select parameters with BO}
            \State Update GP model with $\rho_{\varphi}(\mathbf{e})$ \Comment{Refine GP model with new data}
        \EndFor
        \State Previous search zone: $\mathbf{y} \gets [\rho_{\varphi} (\mathbf{e}_{1}^{(g-1)}), \dots, \rho_{\varphi}(\mathbf{e}_{P}^{(g-1)})]$ 
        \State Current search zone: $\mathbf{z} \gets [\rho_{\varphi}(\mathbf{e}_{1}^{(g)}), \dots, \rho_{\varphi}(\mathbf{e}_{P}^{(g)})]$
        \If{$\min(\mathbf{z}) \leq \min(\mathbf{y})$} \Comment{Stagnation monitor process}
            \State Update $\mathcal{U}_{L}$
        \Else 
            \State $\gamma \gets \gamma + 1$
            \If{$\gamma \geq \delta$}
                \State Shift $\mathcal{U}_{L}$ to an unexplored region of $\mathcal{U}_{G}$
            \EndIf  
        \EndIf
    \end{algorithmic}
\end{algorithm}


\section{Experimental Setups and Case Studies}
\label{evaluation}

We evaluate the proposed method against vanilla BO and CMA-ES on several benchmark problems. For each case study, the methods are exposed to the same uncertainty space $\mathcal{U}_{G}$. \texttt{BEACON} and vanilla BO are subject to simulation budgets of $100$, $200$, $300$, $400$, and $500$ where they perform $150$ tests for each budget. CMA-ES is subject to  $150$ tests per case study and is not restricted to a simulation budget.

For \texttt{BEACON}, we choose several user-defined settings prior to testing that remain consistent across experiments. The local search zone simulation budget $P$ is set to $20$ simulations. For this study, \texttt{BEACON} uses the top quarter $P_{best}=5$ of the environmental parameters to calculate $\mathcal{U}_{L}$. Finally, we set the stagnation constant $\delta=2$ so that \texttt{BEACON} can effectively exploit local areas, but retain resources for wider exploration. Vanilla BO starts with an initial sampling of $20$ parameters, the same as with \texttt{BEACON}. Finally, each generation of CMA-ES performs $20$ simulations from which the top $5$ are used to adapt the covariance matrix.

\subsection{Case Study 1: Mountain Car}
The mountain car environment is an autonomous car situated at the bottom of a valley on a one-dimensional track. The car's objective is to ascend the right hill by employing acceleration in both the left and right directions. The car has two observable states: its position $x$ and its velocity $\dot{x}$. We consider four sources of uncertainty whose ranges are provided in Table \ref{MC Spec}. The uncertain parameters are the initial position $x$, initial velocity $\dot{x}$, the car's maximum velocity $\dot{x}_{max}$, and maximum power magnitude $\rho_{max}$. 

The car is controlled by a policy trained with deep deterministic policy gradient (DDPG) \cite{lillicrap2015continuous}. The controller is subject to two safety specifications simultaneously represented in STL format in Table \ref{MC Spec}. The first specification states that the car's velocity should \emph{always} remain below $0.0735$ when its position is less than $-1.1$ or greater than $0.5$. Second, the car's velocity should remain below $0.055$ \emph{until} it has reached the position $0.1$.

\begin{table*}[t]
    \centering
    \begin{tabular}{ll}
    \begin{minipage}[t]{0.5\textwidth}
        \centering
        \caption{Mountain car specifications and environmental parameters}
        \begin{tabularx}{\linewidth}{p{3cm} X} 
        \toprule
        \textbf{Specification} & \textbf{STL specification}\\
        \midrule
        $\varphi_{1}$ & $\mathbf{G}((x\leq -1.1)\vee x\geq 0.5)\wedge (\dot{x}<0.0735))$\\
        &\\
        $\varphi_{2}$ & $(\dot{x}<0.055)\mathbf{U}(x>0.1)$\\
        \end{tabularx}
        \begin{tabularx}{\linewidth}{p{3cm} p{1.5cm} X} 
        \toprule
        \textbf{Parameter} & \textbf{Symbol} & \textbf{Range}\\
        \midrule
        Position & $x$ & [$-0.6$, $-0.4$]\\
        Velocity & $\dot{x}$ & [$-0.025$, $0.025$]\\
        Maximum velocity & $\dot{x}_{max}$ & [$0.040$, $0.075$]\\
        Power magnitude & $\rho_{max}$ & [$0.0005$, $0.0025$]\\
        \bottomrule
        \end{tabularx}
        \label{MC Spec}
    \end{minipage}
    \hspace{0.02 \linewidth}
    \begin{minipage}[t]{0.48\textwidth}
        \centering
        \caption{Automatic transmission specifications and environmental parameters}
        \begin{tabularx}{\linewidth}{p{3cm} X} 
        \toprule
        \textbf{Specification} & \textbf{STL specification}\\
        \midrule
        $\varphi_{1}$ & $\mathbf{G}(\dot{x}\leq 80)$\\
        $\varphi_{2}$ & $\mathbf{G}(\omega \leq 1400)$\\
        \end{tabularx}
        \begin{tabularx}{\linewidth}{p{3cm} p{1.5cm} X} 
        \toprule
        \textbf{Parameter} & \textbf{Symbol} & \textbf{Range}\\
        \midrule
        Throttle & $\theta_{thr}$ & $[0,100]^{2}$\\
        Brake & $\theta_{brk}$ & $[0,100]^{2}$\\
        \bottomrule
        \end{tabularx}
        \label{AT_Spec}
    \end{minipage}    
    \end{tabular}
\end{table*}

\begin{table*}[b]
\centering
\caption{Neural network specifications and environmental parameters}

\begin{tabularx}{\linewidth}{p{3cm} X} 
\toprule
\textbf{Specification} & \textbf{STL specification}\\
\midrule
$\varphi_{1}$ & $\mathbf{G}(h < 3.9)$\\
$\varphi_{2}$ & $\mathbf{G}_{[0,50]}(\neg(|h-Ref| > 0.005+0.04|Ref|)\Rightarrow \mathbf{F}_{[0,2]}\mathbf{G}_{[0,1]}(\left | h-Ref \right |\leq 0.005+0.04\left | Ref \right |))$\\
\end{tabularx}
\begin{tabularx}{\linewidth}{p{3cm} p{1.5cm} X} 
\toprule
\textbf{Parameter} & \textbf{Symbol} & \textbf{Range}\\
\midrule
Reference & $Ref$ & $[0,3]^{8}$\\
\bottomrule
\end{tabularx}

\label{NN_Table}
\end{table*}
\subsection{Case Study 2: Automatic Transmission}
The automatic transmission environment is a scenario that simulates the speed of a $4$-gear vehicle with an automatic transmission. The simulation has two observable properties: the vehicle's speed $\dot{x}$ and engine speed $\omega$. We consider four sources of uncertainty whose ranges are provided in Table \ref{AT_Spec}. We consider two input signals, the throttle angle $\theta_{thr}$ and brake angle $\theta_{brk}$.

We explore the uncertainty space for a combination of environmental parameters that cause the vehicle to violate one of the following two specifications presented in Table \ref{AT_Spec}. First, the vehicle's speed should \emph{always} remain below $80$mph. Second, the engine speed should \emph{always} remain below $1400$rpm.

\subsection{Case Study 3: Neural Network Controller}
This environment models a magnet levitating above an electromagnet, maintaining a specific reference height. The simulation tracks the height $h$ of the magnet, with the only input being the reference position. 
Thus, the model incorporates eight sources of uncertainty, detailed in Table \ref{NN_Table}.

We evaluate the nonlinear autoregressive moving average (NARMA) neural controller's capability to move the magnet to a reference position by controlling the current \cite{NN_cite}. The neural controller consists of a neural network with nine hidden layers. The controller is subjected to two safety specifications given in STL format in Table \ref{NN_Table}. First, the controller should \emph{always} keep the magnet below $3.9$mm. Second, the magnet should \emph{always} settle to a new reference position within two seconds. This can be reworded as the magnet should \emph{eventually} remain within the specified range of the reference position for one second.

\subsection{Case Study 4: F16-Ground Collision Avoidance}

This environment simulates the F-$16$ control system, with a specific focus on the aircraft’s ground collision avoidance inner-loop controller, modeled by $16$ continuous piece-wise nonlinear differential equations \cite{F16_cite}. Although the F-$16$ environment features a wide range of observable properties, such as roll, pitch, and yaw angles, the primary concern for the falsification problem is the aircraft's altitude.

We falsify the controller against five sources of uncertainty listed in Table \ref{F16_Spec}. The uncertainty parameters consist of the altitude $alt$, initial velocity $\dot{x}$, roll angle $\theta$, pitch angle $\phi$, and yaw angle $\omega$. The controller is subject to the singular safety specification that the aircraft should \emph{always} avoid colliding with the ground during evasive maneuvers.
\begin{table*}[t]
    \centering
    \begin{tabular}{ll}

    \begin{minipage}[t]{0.5\textwidth}
        \centering
        \caption{F16 specifications and environmental \\parameters}
        \begin{tabularx}{\linewidth}{p{2.5cm} X} 
            \toprule
            \textbf{Specification} & \textbf{STL specification}\\
            \midrule
            $\varphi_{1}$ & $\mathbf{G}_{[0,15]}(\text{altitude} > 0)$\\
        \end{tabularx}
        \begin{tabularx}{\linewidth}{p{2.5cm} p{1.5cm} X} 
            \toprule
            \textbf{Parameter} & \textbf{Symbol} & \textbf{Range}\\
            \midrule
            Altitude & $alt$ & [$900$, $4000$]\\
            Velocity & $\dot{x}$ & [$340$, $740$]\\
            Roll angle & $\theta$ & [$0.6283$, $0.8900$]\\
            Pitch angle & $\phi$ & [$-1.6964$, $-1.5707$]\\
            Yaw angle & $\omega$ & [$0.7853$, $1.17809$]\\
            \bottomrule
        \end{tabularx}
        \label{F16_Spec}
    \end{minipage}
    \hspace{0.02 \linewidth}
    \begin{minipage}[t]{0.45\textwidth}
        \centering
        \caption{Air fuel control specification and environmental parameters}
        
        \begin{tabularx}{\linewidth}{p{2.5cm} X} 
            \toprule
            \textbf{Specification} & \textbf{STL specification}\\
            \midrule
            $\varphi_{1}$ & $\mathbf{G}_{[10,30]}(\left | AF-14.7 \right |< \text{tol}\cdot 14.7)$\\
            \end{tabularx}
            \begin{tabularx}{\linewidth}{p{2.5cm} p{1.5cm} X} 
            \toprule
            \textbf{Parameter} & \textbf{Symbol} & \textbf{Range}\\
            \midrule
            Pedal angle & $\theta$ & $[0,61]^{10}$\\
            Engine speed & $\omega$ & [$900$, $1100$]\\
            \bottomrule
        \end{tabularx}
        \label{AFC_Spec}
    \end{minipage}    
    \end{tabular}
\end{table*}
\subsection{Case Study 5: Air Fuel Control}
The air fuel control system model captures the dynamics of fuel regulation, focusing on the air-fuel ratio in response to varying inputs such as throttle angle and engine speed. This model allows us to analyze the behavior of the air-fuel mixture across different operational conditions \cite{jin2014powertrain}. The uncertainty space consists of $11$ parameters whose ranges are given in Table \ref{AFC_Spec}. The system is subjected to one safety parameter provided in Table \ref{AFC_Spec}. The specification states that the air-fuel ratio should \emph{always} remain with $0.7\%$ of the value of $14.7$, otherwise, the system may emit undesirable quantities of noxious fumes.
\begin{figure*}[t]
\centering
  \begin{subfigure}{0.25\textwidth}
    \begin{tikzpicture}[scale=0.5]
       \begin{axis}[
            axis lines=left,
            axis line style={-},
            xlabel={Simulation Budget},
            ylabel={Violation Rate (\%)},
            symbolic x coords={100,200,300,400,500},
            xtick=data,
            xmin=100, xmax=500,
            ymin=0, ymax=100,
            grid=both,
            legend style={at={(1,1)}, anchor=north east,legend columns=-1},
            ytick={0,20,40,60,80,100},
            yticklabel={\pgfmathprintnumber{\tick}\%},
            bar width=0.2cm,
            enlarge x limits=0.15,
        ]
        \addplot[mark=*, only marks, color=blue] coordinates {
            (100,75.3)
            (200,76.5)
            (300,76.3)
            (400,76.1)
            (500,83.2)
        };
        \addplot[mark=*, only marks, color=red] coordinates {
            (100,52.3)
            (200,74.7)
            (300,79.0)
            (400,81.2)
            (500,80.8) 
        };
        \legend{\texttt{BEACON}, BO}
        \end{axis}
    \end{tikzpicture}
    \caption{Mountain Car}
  \end{subfigure}%
  \hspace{.25\textwidth}
  \vspace{0.25cm}
  \begin{subfigure}{0.25\textwidth}
    \begin{tikzpicture}[scale=0.5]
       \begin{axis}[
            axis lines=left,
            axis line style={-},
            xlabel={Simulation Budget},
            ylabel={Violation Rate (\%)},
            symbolic x coords={100,200,300,400,500},
            xtick=data,
            xmin=100, xmax=500,
            ymin=0, ymax=100,
            grid=both,
            legend style={at={(1,1)}, anchor=north east,legend columns=-1},
            ytick={0,20,40,60,80,100},
            yticklabel={\pgfmathprintnumber{\tick}\%},
            bar width=0.2cm,
            enlarge x limits=0.15,
        ]
        
        \addplot[mark=*, only marks, color=blue] coordinates {
            (100,75.2)
            (200,76.5)
            (300,71.8)
            (400,76.0)
            (500,74.6)
        };
        \addplot[mark=*, only marks, color=red] coordinates {
            (100,52.5)
            (200,74.6)
            (300,81.8)
            (400,74.4)
            (500,85.7)
            
        };
        \legend{\texttt{BEACON}, BO}
    
        \end{axis}
    \end{tikzpicture}
    \caption{Automatic Transmission}
  \end{subfigure}
  \hspace{.25\textwidth}
  \vspace{0.25cm}
  \begin{subfigure}{0.25\textwidth}
    \begin{tikzpicture}[scale=0.5]
       \begin{axis}[
            axis lines=left,
            axis line style={-},
            xlabel={Simulation Budget},
            ylabel={Violation Rate (\%)},
            symbolic x coords={100,200,300,400,500},
            xtick=data,
            xmin=100, xmax=500,
            ymin=0, ymax=100,
            grid=both,
            legend style={at={(1,1)}, anchor=north east,legend columns=-1},
            ytick={0,20,40,60,80,100},
            yticklabel={\pgfmathprintnumber{\tick}\%},
            bar width=0.2cm,
            enlarge x limits=0.15,
        ]
        
        \addplot[mark=*, only marks, color=blue] coordinates {
            (100,84.7)
            (200,87.5)
            (300,86.8)
            (400,86.0)
            (500,84.4)
        };
        \addplot[mark=*, only marks, color=red] coordinates {
            (100,79.7)
            (200,78.3)
            (300,77.1)
            (400,77.4)
            (500,75.1)
            
        };
        \legend{\texttt{BEACON}, BO}
    
        \end{axis}
    \end{tikzpicture}
    \caption{Neural Network Controller}
  \end{subfigure}%
  \hspace{.25\textwidth}
  \vspace{0.25cm}
  \begin{subfigure}{0.25\textwidth}
   \begin{tikzpicture}[scale=0.5]
       \begin{axis}[
            axis lines=left,
            axis line style={-},
            xlabel={Simulation Budget},
            ylabel={Violation Rate (\%)},
            symbolic x coords={100,200,300,400,500},
            xtick=data,
            xmin=100, xmax=500,
            ymin=0, ymax=100,
            grid=both,
            legend style={at={(1,1)}, anchor=north east,legend columns=-1},
            ytick={0,20,40,60,80,100},
            yticklabel={\pgfmathprintnumber{\tick}\%},
            bar width=0.2cm,
            enlarge x limits=0.15,
        ]
        
        \addplot[mark=*, only marks, color=blue] coordinates {
            (100,89.9)
            (200,84.2)
            (300,84.2)
            (400,85.5)
            (500,86.5)
        };
        \addplot[mark=*, only marks, color=red] coordinates {
            (100,77.3)
            (200,78.8)
            (300,78.7)
            (400,80.1)
            (500,81.4)
            
        };
        \legend{\texttt{BEACON}, BO}
    
        \end{axis}
    \end{tikzpicture}
    \caption{F-16 Control System}
  \end{subfigure}
  \hspace{.25\textwidth}
  \vspace{0.25cm}
  \begin{subfigure}{0.25\textwidth}
   \begin{tikzpicture}[scale=0.5]
       \begin{axis}[
            axis lines=left,
            axis line style={-},
            xlabel={Simulation Budget},
            ylabel={Violation Rate (\%)},
            symbolic x coords={100,200,300,400,500},
            xtick=data,
            xmin=100, xmax=500,
            ymin=0, ymax=100,
            grid=both,
            legend style={at={(1,1)}, anchor=north east,legend columns=-1},
            ytick={0,20,40,60,80,100},
            yticklabel={\pgfmathprintnumber{\tick}\%},
            bar width=0.2cm,
            enlarge x limits=0.15,
        ]
        
        \addplot[mark=*, only marks, color=blue] coordinates {
            (100,15.4)
            (200,15.5)
            (300,15.4)
            (400,15.4)
            (500,15.5)
        };
        \addplot[mark=*, only marks, color=red] coordinates {
            (100,15.5)
            (200,15.1)
            (300,15.0)
            (400,15.6)
            (500,15.0)
            
        };
        \legend{\texttt{BEACON}, BO}
    
        \end{axis}
    \end{tikzpicture}
    \caption{Air Fuel Control}
    \label{AFC_Fig1}
  \end{subfigure}%
  \hspace{.25\textwidth}
  \vspace{0.25cm}
  \begin{subfigure}{0.25\textwidth}
   \begin{tikzpicture}[scale=0.5]
       \begin{axis}[
            axis lines=left,
            axis line style={-},
            xlabel={Simulation Budget},
            ylabel={Violation Rate (\%)},
            symbolic x coords={100,200,300,400,500},
            xtick=data,
            xmin=100, xmax=500,
            ymin=10, ymax=20,
            grid=both,
            legend style={at={(1,1)}, anchor=north east,legend columns=-1},
            ytick={10,12,14,16,18,20},
            yticklabel={\pgfmathprintnumber{\tick}\%},
            bar width=0.2cm,
            enlarge x limits=0.15,
        ]
        
        \addplot[mark=*, only marks, color=blue] coordinates {
            (100,15.4)
            (200,15.5)
            (300,15.4)
            (400,15.4)
            (500,15.5)
        };
        \addplot[mark=*, only marks, color=red] coordinates {
            (100,15.5)
            (200,15.1)
            (300,15.0)
            (400,15.6)
            (500,15.0)
            
        };
        \legend{\texttt{BEACON}, BO}
    
        \end{axis}
    \end{tikzpicture}
    \caption{Air Fuel Control (Zoomed)}
    \label{AFC_Fig2}
  \end{subfigure}
  \caption{The illustration of violation rate vs. simulation budget for the mountain car, automatic transmission, neural network, F16, and air fuel control case studies. In these plots, \texttt{BEACON}'s results are presented in blue, and BO's results are presented in red. \texttt{BEACON} performs better than BO at lower simulation budgets in the cases of mountain car and automatic transmission. In the air fuel control case study, \texttt{BEACON} and BO performed similarly across each budget. In the neural network and F-$16$ environments, \texttt{BEACON} achieves higher violation rates for each budget than BO.}
  \label{Viol_Rate_Plot}
\end{figure*}
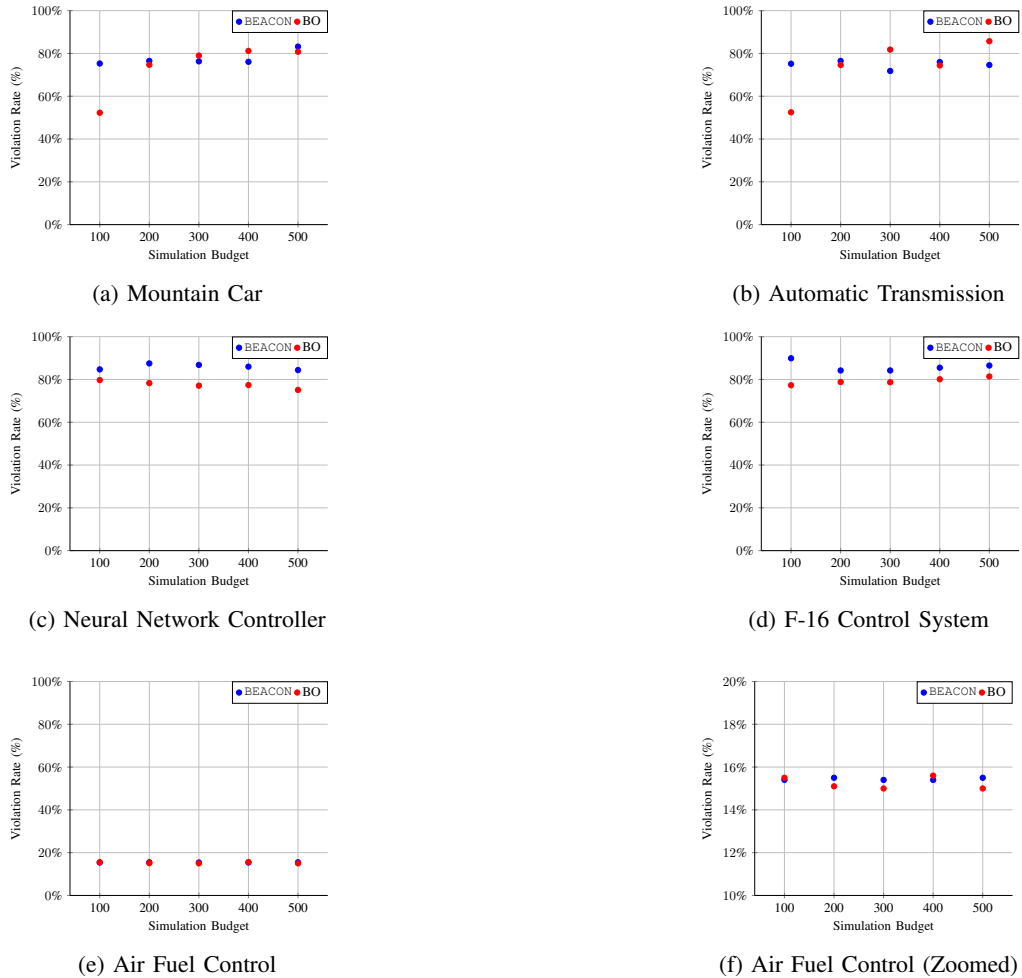

\begin{table*}[b]
    \centering
    \caption{Results for each case study found with \texttt{BEACON}, BO, and CMA-ES.}
    \begin{tabularx}{\textwidth}{X X *{6}{X}}
    \toprule
    \textbf{Case} & \textbf{Simulation} & \multicolumn{2}{c}{\texttt{\textbf{BEACON}}} & \multicolumn{2}{c}{\textbf{BO}} & \multicolumn{2}{c}{\textbf{CMA-ES}} \\
  \textbf{Study} & \textbf{Budget} & \textbf{Violation Rate} & \textbf{Sims/ $\text{Violation}$} & \textbf{Violation Rate} & \textbf{Sims/ $\text{Violation}$} & \textbf{Violation Rate} & \textbf{Sims}\\
    \midrule
    \multirow{5}{*}{$\boldsymbol{\varphi}_{\mathbf{1,2}}^{\textbf{MC}}$} & 100 & \textbf{75.3\%} & 1.33 & 52.3\% & 1.90  & \multirow{5}{*}{86.9\%} & \multirow{5}{*}{4788}\\
    & 200  & \textbf{76.5\%} & 1.31 & 74.7\% & 1.33 &  &\\
    & 300 & 76.3\% & 1.31 & \textbf{79.0\%} & 1.27 & &\\
    & 400 & 76.1\% & 1.31 & \textbf{81.2\%} & 1.23 & &\\
    & 500 & \hl{\textbf{83.2\%}} & 1.21 & 80.8\% & 1.24 & &\\
    \midrule
    \multirow{5}{*}{$\boldsymbol{\varphi}_{\mathbf{1,2}}^{\textbf{AT}}$} & 100 &  \textbf{75.2\%} & 1.33 & 52.5\% & 1.90 & \multirow{5}{*}{83.7\%} & \multirow{5}{*}{4744}\\
    & 200 & \textbf{76.5\%} & 1.32 & 74.6\% & 1.34 &  & \\
    & 300 & 71.8\% & 1.39 & \textbf{81.8\%} & 1.22 & &\\
    & 400 & \textbf{76.0\%} & 1.32 & 74.4\% & 1.34 & &\\
    & 500 & 74.6\% & 1.34 & \hl{\textbf{85.7\%}} & 1.16 & &\\
    \midrule
    \multirow{5}{*}{$\boldsymbol{\varphi}_{\mathbf{1,2}}^{\textbf{NN}}$} & 100 & \textbf{84.7\%} & 1.18 & 79.7\% & 1.26 & \multirow{5}{*}{91.9\%} & \multirow{5}{*}{1157}\\
    & 200 & \hl{\textbf{87.5\%}} & 1.14 & 78.3\% & 1.28 &  & \\
    & 300 & \textbf{86.8\%} & 1.15 & 77.1\% & 1.30 & &\\
    & 400 & \textbf{86.0\%} & 1.16 & 77.4\% & 1.29 & &\\
    & 500 & \textbf{84.4\%} & 1.19 & 75.1\% & 1.33 & &\\
    \midrule
    \multirow{5}{*}{$\boldsymbol{\varphi}_{\mathbf{1}}^{\textbf{F16}}$} & 100 & \hl{\textbf{89.9\%}} & 1.11 & 77.3\% & 1.29 & \multirow{5}{*}{53.9\%} & \multirow{5}{*}{6442}\\
    & 200 & \textbf{84.2\%} & 1.19 & 78.8\% & 1.26 &  & \\
    & 300 & \textbf{84.2\%} & 1.19 & 78.7\% & 1.27 & &\\
    & 400 & \textbf{85.5\%} & 1.17 & 80.1\% & 1.25 & &\\
    & 500 & \textbf{86.5\%} & 1.16 & 81.4\% & 1.23 & &\\
    \midrule
    \multirow{5}{*}{$\boldsymbol{\varphi}_{\mathbf{1,2}}^{\textbf{AFC}}$} & 100 & 15.4\% & 6.48 & \textbf{15.5\%} & 6.47 & \multirow{5}{*}{16.9\%} & \multirow{5}{*}{14363}\\
    & 200 & \textbf{15.5\%} & 6.47 & 15.1 & 6.61 & & \\
    & 300 & \textbf{15.4\%} & 6.49 & 15.0\% & 6.66 & &\\
    & 400 & 15.4\% & 6.49 & \hl{\textbf{15.6\%}} & 6.43 & &\\
    & 500 & \textbf{15.5\%} & 6.45 & 15.0\% & 6.66 & &\\
    \bottomrule
    \end{tabularx}
    \label{Results_Table}
\end{table*}

\section{Results and Discussion}
\label{Results} We present the violation rates and simulations required to locate a violation for each experiment from \texttt{BEACON}, BO, and CMA-ES in Table \ref{Results_Table}. The highlighted numbers indicate the highest violation rate achieved within a case study between \texttt{BEACON} and BO. Numbers in bold represent the higher violation rate obtained in each experiment. The results for \texttt{BEACON} and BO are visualized in Figure \ref{Viol_Rate_Plot} for each case study.

In our initial case study, the mountain car scenario, \texttt{BEACON} consistently outperforms BO. \texttt{BEACON} achieves an average violation rate of $77.5\%$  compared to $73.7\%$ for BO. In particular, \texttt{BEACON} achieves a higher violation rate in three of the five experiments which have simulation budgets of $100$, $200$, and $500$. CMA-ES, while exhibiting a violation rate of $86.9$\%, is accompanied by a significant resource demand of $4788$ simulations, in contrast to \texttt{BEACON}'s highest performance at $500$ simulations with a violation rate of $83.2$\%. \texttt{BEACON}'s results are within $3.7$\% of those achieved by CMA-ES, all while utilizing only a tenth of the resource budget. Similar trends are recorded in the automatic transmission environment. Here, \texttt{BEACON} maintains an average violation rate of $74.8$\%, surpassing BO's $73.8$\%. In this case study, \texttt{BEACON} secures higher violation rates in three of the experiments given $100$, $200$, and $400$ simulations. CMA-ES, with an average violation rate of $83.7$\%, comes with a cost of $4744$ simulations. In contrast, \texttt{BEACON} reaches its highest rate with $200$ simulations at $76.5$\%. \texttt{BEACON}, requiring only $4$\% of the simulations to achieve within $7.2$\% of CMA-ES's violation rate.

\texttt{BEACON} demonstrates its second-highest performance in the neural network environment, consistently outperforming BO. \texttt{BEACON} achieves an average violation rate of $85.9$\% compared to BO's $77.5$\%. Although CMA-ES reaches its peak performance with a violation rate of $91.9$\%, this comes at the expense of an average of $1157$ simulations. \texttt{BEACON}, on the other hand, achieves a rate of $87.5$\% with $200$ simulations, delivering results within $4.4$\% of CMA-ES with only a fifth of the resources.
\begin{figure}
\pgfplotsset{compat=1.11,
    /pgfplots/ybar legend/.style={
    /pgfplots/legend image code/.code={%
       \draw[##1,/tikz/.cd,yshift=-0.25em]
        (0cm,0cm) rectangle (3pt,0.8em);},
   },
}
    \centering
\begin{tikzpicture}
    \begin{axis}[
    axis lines=left,
    axis line style={-},
    width=3in,
    height=2in,
    grid=none,
    ybar,
        xlabel={Case Study},
        ylabel={Violation Rate (\%)},
        symbolic x coords={MC, AT, NN, F16, AFC}, 
        xtick=data, 
        xmin=MC, xmax=AFC, 
        ymin=0, ymax=120, 
        grid=none,
        legend style={at= {(1,1)},
      anchor=north east,legend columns=-1},
        ytick={0,20,40,60,80,100}, 
        yticklabel={\pgfmathprintnumber{\tick}\%}, 
        bar width=0.2cm,
        enlarge x limits=0.15,
    ]
    
    \addplot[ybar, fill=blue] coordinates {
        (MC, 77.48)
        (AT, 74.82)
        (NN, 85.88)
        (F16, 86.06)
        (AFC, 15.44)
        };
    \addplot[ybar, fill=red] coordinates {
        (MC,73.6)
        (AT, 73.8)
        (NN, 77.52)
        (F16,79.06)
        (AFC, 15.24)
        };
    \addplot[ybar, fill=orange] coordinates {
    (MC,86.9)
    (AT,83.7 )
    (NN, 91.9)
    (F16,53.9)
    (AFC, 16.9)
    };
    \legend{\texttt{BEACON}, BO, CMA-ES}

    \end{axis}
\end{tikzpicture}

\caption{Comparative analysis of violation rates across case studies.}
\label{Mean Violation Plot}
\end{figure}
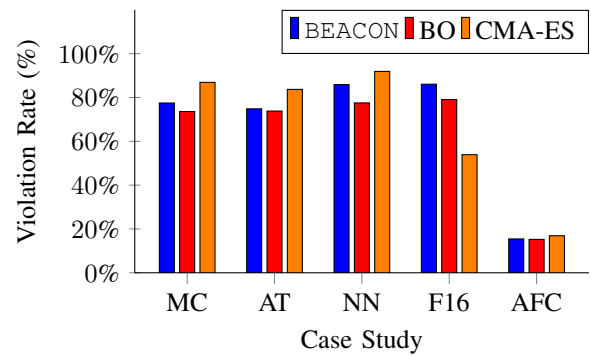
In the F-$16$ case study, \texttt{BEACON} shines, achieving an average violation rate of $86.1$\%. This rate is $7$\% higher than that of BO's result and $36$\% higher than CMA-ES's $53.9$\%. Notably, \texttt{BEACON} outperforms BO and CMA-ES in each experiment. \texttt{BEACON}'s lowest rate in this scenario, $84.2$\%, is achieved with $200/300$ simulations, while BO's highest rate is $81.4$\% at $500$ simulations, falling $2.8$\% short of \texttt{BEACON}'s performance with a $1.5$ times larger budget.

In the final case study, air fuel control, all three methods obtain similar rates. On average, \texttt{BEACON} achieves a rate of $15.44$\%, compared to $15.24$\% with BO, and $16.9$\% with CMA-ES. \texttt{BEACON} outperforms BO in three out of five experiments when given $200$, $300$, and $500$ simulation budgets. CMA-ES requires $14363$ simulations on average to achieve its $16.9$\% violation rate. \texttt{BEACON} performs within $1.4$\% of CMA-ES with $3.5$\% of the simulations.

Across the five experiments for each case study, \texttt{BEACON} consistently outperforms BO by an average margin ranging from $0.2$\% to $8.3$\% as depicted in Figure \ref{Mean Violation Plot}. Table \ref{Resource_Table} provides further insight for our comparison by presenting the highest violation rate achieved with each method along with the required simulation budgets (in parentheses). In Table \ref{Resource_Table}, green highlighting indicates the highest violation rate achieved by either \texttt{BEACON} or BO in a given case study, yellow highlights denote instances where CMA-ES achieved the highest violation rate, and red highlights signify the largest simulation budget required by the three methods. In three of the case studies, CMA-ES achieves the highest violation rate. However, in all cases, this approach requires far more resources to achieve its results compared to \texttt{BEACON} which was discussed in each case study. \texttt{BEACON} achieves the highest rate out of the three methods in the F-$16$ environment, and higher rates than BO in mountain car and neural network. From the data, we can observe that \texttt{BEACON} tends to achieve its highest rates with $200$ or fewer simulations compared to BO whose highest rates occur mostly from $400-500$ simulation budgets.

Several key conclusions can be drawn from these results. \texttt{BEACON} excels in situations where locating violations can prove challenging, such as in the neural network and F-$16$ environments. Additionally, \texttt{BEACON} operates more efficiently at lower simulation budgets, primarily due to its ability to perform multiple uncertainty space searches before exploring unknown regions. In contrast, BO tends to shift from exploration to exploitation as it acquires information. Overall, \texttt{BEACON} performs on par with or better than BO, depending on the circumstance, and is capable of achieving similar results to CMA-ES with significantly fewer resources.

\noindent{\textbf{Limitations.}} The \texttt{BEACON} framework, while promising, has certain limitations that should be acknowledged and addressed in future research. One limitation lies in the assumption of continuous and smooth robustness functions. In real-world systems, the robustness function may exhibit discontinuities or non-smooth behavior, which can impact the effectiveness of the GP model and the overall search process. Discontinuities can arise from abrupt changes in system dynamics or from the discrete nature of certain environmental parameters. Another limitation of \texttt{BEACON} is its scalability to high-dimensional search spaces. As the number of dimensions increases, the volume of the search space grows exponentially, making it challenging to efficiently explore and identify counterexamples. The curse of dimensionality can hinder the performance of \texttt{BEACON}, particularly in systems with a large number of environmental parameters. This limitation calls for the development of advanced sampling techniques, dimensionality reduction methods, and efficient surrogate models that can handle high-dimensional spaces. 

\begin{table}[t]
 \caption{Comparison of violation rates and simulation counts across different methodologies for each case study. The table showcases the highest violation rates attained by \texttt{BEACON}, BO, and CMA-ES. Parentheses indicate the number of simulations conducted to reach the noted violation rate. Green highlights denote instances where \texttt{BEACON} or BO demonstrates superior performance within the comparison, whereas yellow highlights emphasize case studies where CMA-ES outperforms. Orange highlights signify the scenarios demanding the highest simulation effort to achieve the reported outcomes.}

\begin{tabularx}{\linewidth}{p{1.2cm}|X|X|X} 
\toprule
\textbf{Case Study} & \texttt{\textbf{BEACON}} & \textbf{BO} & \textbf{CMA-ES}\\
\midrule
$\boldsymbol{\varphi}_{\mathbf{1,2}}^{\textbf{MC}}$ & \hlc[green]{83.2\%} (500) & 81.2\% (400) & \hl{86.9\%} (\hlc[orange]{4788})\\
\midrule
$\boldsymbol{\varphi}_{\mathbf{1,2}}^{\textbf{AT}}$ & 76.5\% (200) & \hlc[green]{85.7\%} (500) & 83.7\% (\hlc[orange]{4744})\\
\midrule
$\boldsymbol{\varphi}_{\mathbf{1,2}}^{\textbf{NN}}$ & \hlc[green]{87.5\%} (200) & 79.7\% (100) & \hl{91.9\%} (\hlc[orange]{1157})\\
\midrule
$\boldsymbol{\varphi}_{\mathbf{1}}^{\textbf{F16}}$ & \hlc[green]{89.9\%}(100) & 81.4\% (500) & 53.9\% (\hlc[orange]{6442})\\
\midrule
$\boldsymbol{\varphi}_{\mathbf{1,2}}^{\textbf{AFC}}$ & 15.5\%(500) & \hlc[green]{15.6\%}(400) & \hl{16.9\%} (\hlc[orange]{14363})\\
\bottomrule
\end{tabularx}
\label{Resource_Table}
\end{table}
\section{Conclusions}
\label{Conclusion}

In this work, we have proposed \texttt{BEACON}, a novel hybrid falsification framework that integrates Bayesian optimization and covariance matrix adaptation evolutionary strategy, aiming to enhance the efficiency of safety violation detection in control systems. \texttt{BEACON} segments the global parameter space into localized search zones, enabling the generation of accurate surrogate models to guide the selection of environmental parameters more effectively. Through comprehensive evaluation across diverse case studies, \texttt{BEACON} has demonstrated its capability to not only match but in certain instances surpass the efficacy of its constituent methodologies in identifying counterexamples.

There are several exciting avenues for future research and development that can further enhance the capabilities and performance of the \texttt{BEACON}. One crucial direction is to investigate the integration of dynamic parameter ranges from the CMA-ES component into the BO process. Currently, the BO assumes fixed uncertainty spaces for each parameter, which may not fully capture the evolving nature of the search space as it is adapted by CMA-ES. By incorporating techniques to update the surrogate model and acquisition function based on the dynamic parameter ranges, \texttt{BEACON} could more accurately model the search space and make informed decisions during the falsification process. This integration has the potential to significantly improve the efficiency and effectiveness of the framework in discovering counterexamples. Another promising avenue is to explore the application of \texttt{BEACON} to a wider range of complex scenarios and safety specifications. Conducting extensive experiments with diverse and challenging falsification tasks will provide valuable insights into the scalability, robustness, and generalizability of the framework. By considering a broader spectrum of specifications and system complexities, we can assess the performance of \texttt{BEACON} in real-world settings and identify areas for further improvement.

\bibliographystyle{IEEEtran}
\bibliography{main}

\end{document}